\documentclass[runningheads]{llncs}
\usepackage{makeidx}
\usepackage[utf8]{inputenc}
\usepackage{enumitem} 
\usepackage{listings} 
\usepackage{color} 
\usepackage[hyphens]{url}
\PassOptionsToPackage{hyphens}{url}
\usepackage{hyperref} 
\usepackage{graphicx} 
\usepackage{multirow} 
\usepackage{wasysym}

\usepackage{datetime}

\newtheorem{requirement}{Requirement}

\definecolor{dkgreen}{rgb}{0,0.6,0}
\definecolor{gray}{rgb}{0.5,0.5,0.5}
\definecolor{mauve}{rgb}{0.58,0,0.82}
\definecolor{lightgray}{rgb}{0.8,0.8,0.8}
\definecolor{darkblue}{rgb}{0,0,.5}

\lstdefinelanguage{JavaScript}{
  keywords={break, case, catch, continue, debugger, default, delete, do, else, finally, for, function, if, in, instanceof, new, return, switch, this, throw, try, typeof, var, void, while, with, encodeURIComponent},
  morecomment=[l]{//},
  morecomment=[s]{/*}{*/},
  morestring=[b]',
  morestring=[b]",
  sensitive=true
}

\hypersetup{
	colorlinks=true,
	breaklinks=true,
	linkcolor=darkblue,
	menucolor=darkblue,
	urlcolor=darkblue,
	citecolor=darkblue,
	pagecolor=darkblue,
	filecolor=darkblue
}

\newdateformat{mydate}{\monthname[\THEMONTH] \number\day, \THEYEAR}

\newcommand*{\plogo}{\includegraphics[width=0.7\textwidth, trim=3cm 14cm 3cm 9cm]{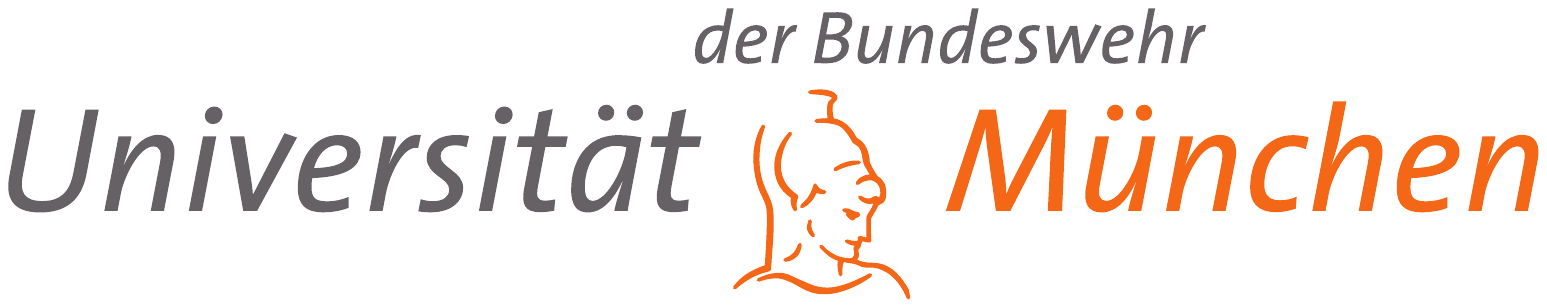}} 

\newcommand*{\titleGM}{\begingroup 
\hbox{ 
\hspace*{0.05\textwidth} 
\rule{1pt}{\textheight} 
\hspace*{0.05\textwidth} 
\parbox[b]{0.9\textwidth}{ 

{\noindent
\huge\bfseries RDF Translator: \\[0.5\baselineskip]
A RESTful Multi-Format \\[0.5\baselineskip]
Data Converter for \\[0.5\baselineskip]
the Semantic Web}\\[2\baselineskip] 
{
\Large Technical Report\\[0.3\baselineskip]
\large\textit{TR-2013-1}\\[0.8\baselineskip]
Version: July 25, 2013}\\[4\baselineskip] 
{
\Large Alex Stolz, \texttt{alex.stolz@unibw.de}\\[0.4\baselineskip]
\Large Bene Rodriguez-Castro, \texttt{bene.rodriguez@unibw.de}\\[0.4\baselineskip]
\Large Martin Hepp, \texttt{martin.hepp@unibw.de}
}\\[2\baselineskip] 

{\Large E-Business and Web Science Research Group,\\[0.2\baselineskip]
Universität der Bundeswehr München\\[0.4\baselineskip]
\large Werner-Heisenberg-Weg 39, D-85579 Neubiberg, Germany}\\[2\baselineskip]

\vspace{1cm} 
{\plogo} 
}}
\endgroup}

\begin{document}

\begin{titlepage}
\titleGM 
\end{titlepage}

\title{RDF Translator: A RESTful Multi-Format Data Converter for the Semantic Web}
\titlerunning{Technical Report - RDF Translator}
\author{Alex Stolz, Bene Rodriguez-Castro, and Martin Hepp}
\authorrunning{A. Stolz, B. Rodriguez-Castro, and M. Hepp}
\institute{
	E-Business and Web Science Research Group, Universität der Bundeswehr München\\
	Werner-Heisenberg-Weg 39, D-85579 Neubiberg, Germany\\
	\email{\{alex.stolz,bene.rodriguez,martin.hepp\}@unibw.de}
}
\maketitle

\pagestyle{headings}
\pagenumbering{arabic}

\begin{abstract}
The interdisciplinary nature of the Semantic Web and the many projects put forward by the community led to a large number of widely accepted serialization formats for RDF. Most of these RDF syntaxes have been developed out of a necessity to serve specific purposes better than existing ones, e.g. RDFa was proposed as an extension to HTML for embedding non-intrusive RDF statements in human-readable documents. Nonetheless, the RDF serialization formats are generally transducible among themselves given that they are commonly based on the RDF model. 
In this paper, we present (1) a RESTful Web service based on the HTTP protocol that translates between different serializations.
In addition to its core functionality, our proposed solution provides (2) features to accommodate frequent needs of Semantic Web developers, namely
a straightforward user interface with copy-to-clipboard functionality,
syntax highlighting,
persistent URI links for easy sharing,
cool URI patterns, and
content negotiation using respective HTTP headers.
We demonstrate the benefit of our converter by presenting two use cases.
\end{abstract}



\section{Introduction}

In the last five years, Semantic Web developers were confronted with an increasing number of alternative syntaxes for publishing RDF content on the Web.
While Uniform Resource Identifiers (URIs) and the Resource Description Framework (RDF) data model \cite{Manola2004} could be established as de-facto components of the Semantic Web, over time there have emerged a number of competing RDF serialization formats (also syntaxes, data formats). While in the early days of the Semantic Web there only existed XML as the standard serialization format for RDF (as depicted in the Semantic Web stack\footnote{\url{http://www.w3.org/2000/Talks/1206-xml2k-tbl/slide10-0.html}}), the most prominent syntaxes for RDF nowadays are
RDF/XML,
N-Triples,
Notation 3 (N3) that embraces Turtle and N-Triples,
RDF in attributes (RDFa), and
JSON.
RDFa is currently the most found syntax on the Web for publishing semantic content in Web pages \cite{Mika2011,Muehleisen2012}.
Microdata constitutes an alternative syntax for embedding structured data in HTML, promoted by search engine operators like Google, Microsoft, Yandex and Yahoo! in the context of \textit{schema.org}\footnote{\url{http://schema.org/}}, a Web vocabulary aimed to be commonly understood by all search engines. The Microdata format was initially designed under the umbrella of the Web Hypertext Application Technology Working Group (WHATWG\footnote{\url{http://www.whatwg.org/}}) and later blended with work at W3C. Unlike RDFa, it relies on a frame-based data structure (representing information as a tree) 
with nested groups of name-value pairs (``items'', setting the context)~\cite{Hickson2012} and thus it is not fully compatible with the more flexible graph-based RDF data model. To give an example, in Microdata it is still not possible to express datatype information like in RDFa. In a blog post in 2011 \cite{Sporny2011}, Manu Sporny summarized the key differences and commonalities between RDFa and Microdata (and Microformats). However, there are ongoing efforts to make Microdata and RDFa compatible with each other, carried on by the technical architecture group (TAG) at W3C. At the time of writing this paper, large parts of Microdata could already be translated seamlessly into RDF syntaxes, and vice versa.

Semantic Web developers may choose between alternative serialization formats based on personal taste or intended purpose. While RDF/XML usage traces back to the beginnings of RDF and is, similar to N-Triples, preferably used in projects with information exchange between systems and/or where large datasets are handled, RDFa and JSON (or JSON-LD) are designed for embedding RDF in HTML markup and for easing the consumption of RDF by Web applications, respectively. Turtle and Notation 3 \cite{Berners-Lee2011} represent abbreviated syntaxes appropriate for human readability by extending N-Triples with support for compact URIs and adding syntactical sugar to shorten the fairly verbose N-Triples syntax.

It is no surprise that the above-mentioned wealth of syntaxes for the Semantic Web can pose serious limitations on interoperability between tools and at the same moment be burdensome for Semantic Web developers that have to deal with various syntactical variants. A Semantic Web developer not acquainted with RDFa, for example, may not wish to be confronted with RDF embedded in HTML using RDFa. He would rather like to read RDF data in his preferred syntax. Similarly, a JavaScript library that a Web site relies on might not support parsing RDFa from other Web pages, albeit being capable of processing JSON-LD.

Obviously, there is much value in translating between those different serialization formats. What is missing is a comprehensive online converter that can fulfill this task, and at the same time permits:
\begin{itemize}
\item developers to quickly test and check annotations that are encoded in less convenient formats (e.g. RDFa in HTML translated to N3),
\item developers to first model in a human-friendlier format and afterwards convert to the target format (e.g. modeling in N3 and publication as RDF/XML),
\item applications to better interoperate among each other.
\end{itemize}

In this paper, we present a Web service to convert between the most prominent serialization formats available on the Web. The service takes advantage of a REST-style architecture, thus being scalable, stateless and supporting HTTP GET and HTTP POST methods for translating documents or textual input. Our main motivation to develop this tool was the lack of free online services that provide quick and hassle-free conversions between the most important syntaxes for the Semantic Web. There exist such services, however, either they are not comprehensive enough by providing only conversions between a syntax pair, or their architecture is not fully REST-compliant, i.e. they may lack suitable identifiers for representing resources, not allow for content negotiation initiated by the client, or similar. Moreover, they are often technically inspired and miss out the human aspect of the development process, such as supporting syntax highlighting for better readability.
Also, since so far there exist no comprehensive, RDFLib-based conversion services, this is the only service suitable for developers that consider to rely their applications on RDFLib and want to have a sandboxed playground for testing purposes.

The main contributions of our approach are summarized as follows:
(1) We comprise two-way conversions for seven popular structured data formats (including Microdata) which should meet most developers' needs;
(2) we provide a user-friendly conversion service, that lowers the barriers because being available online for free and including useful design choices like keyboard shortcuts, easy copy-and-paste functionality, etc.;
(3) our converter will also report syntax errors and is thus appropriate for quick syntax validation; and finally,
(4) due to the REST API and HTTP content negotiation, heterogeneous sources that rely on different data formats can be integrated effortlessly.

The rest of the paper is structured as follows:
Section \ref{sec:relatedwork} summarizes existing tools for converting between RDF formats on the Semantic Web;
in Section \ref{sec:approach}, we introduce our conversion Web service, and in
Section \ref{sec:usecases} we present common use cases to highlight some of the benefits of our online converter.
We finally conclude our work in Section \ref{sec:conclusions}.


\section{Related Tools}
\label{sec:relatedwork}

There exist a number of online tools for the conversion between serialization formats for RDF.
Some of these tools have their origins in offline converters that are made available online as Web services, while others are pure Web services to offer conversions between various Web data formats.
Some of the available Web services are limited to transformations between specific syntaxes. Consequently, we will only be able to report the most popular ones of this kind.

\textbf{Any23}\footnote{\url{http://any23.apache.org/}} (``Anything to Triples'') is a framework that was initially proposed as a powerful Java library and command-line tool for parsing and serializing a variety of Web document formats, used by the Sindice Semantic Web search engine. Soon it was also featured as a public REST-style Web service\footnote{\url{http://any23.org/}} with form-based submission mode, support for cool URIs, GET and POST requests, content negotiation, and advanced error reporting with proper HTTP status codes.
\textbf{Triplr}\footnote{\url{http://triplr.org/}} (``stuff in, triples out'') was one of the earliest RDF format converters for the Semantic Web announced in 2007. It is based on \textit{Raptor}, an RDF syntax library that is part of the Redland \textit{librdf} package written in C. At the time Triplr was designed, it was already capable of guessing the input format of the supplied data source, which is a very useful and much-needed functionality for online converters. Nevertheless, the service is provided as a raw REST service without any HTML-based input form fields that would support users in composing the REST URIs.
\textbf{RDF Distiller}\footnote{\url{http://rdf.greggkellogg.net/distiller}} emerged from a similar motivation as Triplr, namely for the provision of a \textit{Ruby Gem} for Semantic Web development with the Ruby programming language. It has also been deployed as a public Web service and is updated frequently as soon as new features arrive. RDF Distiller provides transformation support for a wealth of RDF syntaxes.

\textbf{Omnidator}\footnote{\url{http://omnidator.appspot.com/}} (``Omnipotent Data Translator'') is a pure Web service intended as a data format translator between formats containing \textit{schema.org} terms into other syntaxes. Currently, it is limited to process \textit{schema.org} data in CSV and Microdata. It can turn input data into JSON, RDF/XML, or Turtle.

In parallel to the tools and services presented so far, there exist Web services solely focusing on conversions between specific data formats, e.g. transforming RDFa into RDF/XML or Notation 3 (N3), but not vice versa.
\textbf{RDFa 1.1 Distiller and Parser}\footnote{\url{http://www.w3.org/2012/pyRdfa/}} is the official RDFa parsing service maintained by W3C that translates to RDF/XML, Turtle, N-Triples and JSON-LD. The service relies on the Python RDF package \textit{RDFLib} and related libraries for parsing RDFa (\textit{pyRDFa}) and for serializing to the various output formats. Besides URI and textual input it allows to upload local files with RDFa embedded in various syntaxes like HTML5, XHTML, XML, Atom, and SVG.
Similar, but not as powerful services for converting RDFa to other Web formats are provided by the \textbf{RDFa Online Parser}\footnote{\url{http://rdf-in-html.appspot.com/}} and \textbf{RDFa Lite 2 RDF Extractor}\footnote{\url{http://getschema.org/rdfaliteextractor/about}}. The latter is powered by \textit{node.js}. A similar RDF extractor also exists for the Microdata syntax\footnote{\url{http://getschema.org/microdataextractor/about}}.

None of the Web services presented so far support the conversion from traditional RDF syntaxes to RDF embedded in Web document formats like RDFa and Microdata in HTML. First attempts to do this resulted in the proposal of the \textit{Snippet Style} approach  for RDFa \cite{Hepp2009:rdf2rdfa}. In this context, an online converter termed \textbf{RDF2RDFa}\footnote{\url{http://www.ebusiness-unibw.org/tools/rdf2rdfa/}} was presented to turn RDF/XML content into RDFa. Later followed a Microdata variant of that service, namely \textbf{RDF2Microdata}\footnote{\url{http://www.ebusiness-unibw.org/tools/rdf2microdata/}}. To the best of our knowledge, these services were the first to generate RDFa or Microdata snippets for embedding RDF content in HTML.
A similar functionality is provided by the RDFa serializer plugin in the PHP ARC2 library, showcased online\footnote{\url{http://zazi.smiy.org/rdfaparser.html}} as a Web service.

A powerful, yet outdated and apparently shutdown service was \textbf{Talis Morph}\footnote{\url{http://www.w3.org/2001/sw/wiki/Morph}} that was able to transform between different Semantic Web formats.
Furthermore, there are some RDF validators that provide in addition to their core functionality, i.e. the validation, also conversion functionality. \textbf{\textit{rdf:about}}\footnote{\url{http://rdfabout.com/demo/validator/}}, for example, is providing an input text area to convert between RDF/XML and Notation 3 (including Turtle and N-Triples).

%
%
%


\section{RDF Translator} 
\label{sec:approach}

In this section, we present \textit{RDF Translator}\footnote{\url{http://rdf-translator.appspot.com/}}, a Web service for bidirectionally converting between various RDF data formats, namely RDFa, Microdata, RDF/XML, Notation 3, N-Triples, RDF/JSON, and JSON-LD. Unlike most of the tools presented in Section \ref{sec:relatedwork}, our online converter is a comprehensive and convenient Web service with having Semantic Web developers in mind.

Subsequently we outline the service architecture, followed by the description of the user interface and what features there are to ensure the developer-friendliness of our Web service. This is afterwards complemented by the specification of the underlying REST API and the related compliance with best practices in Web technology design, i.e. cool URIs, content negotiation, and CORS support. We further address how we solved prefix resolution for better readability of the resulting output.

\subsection{Service Architecture}

\begin{figure}[t]
\center
\includegraphics[width=\textwidth]{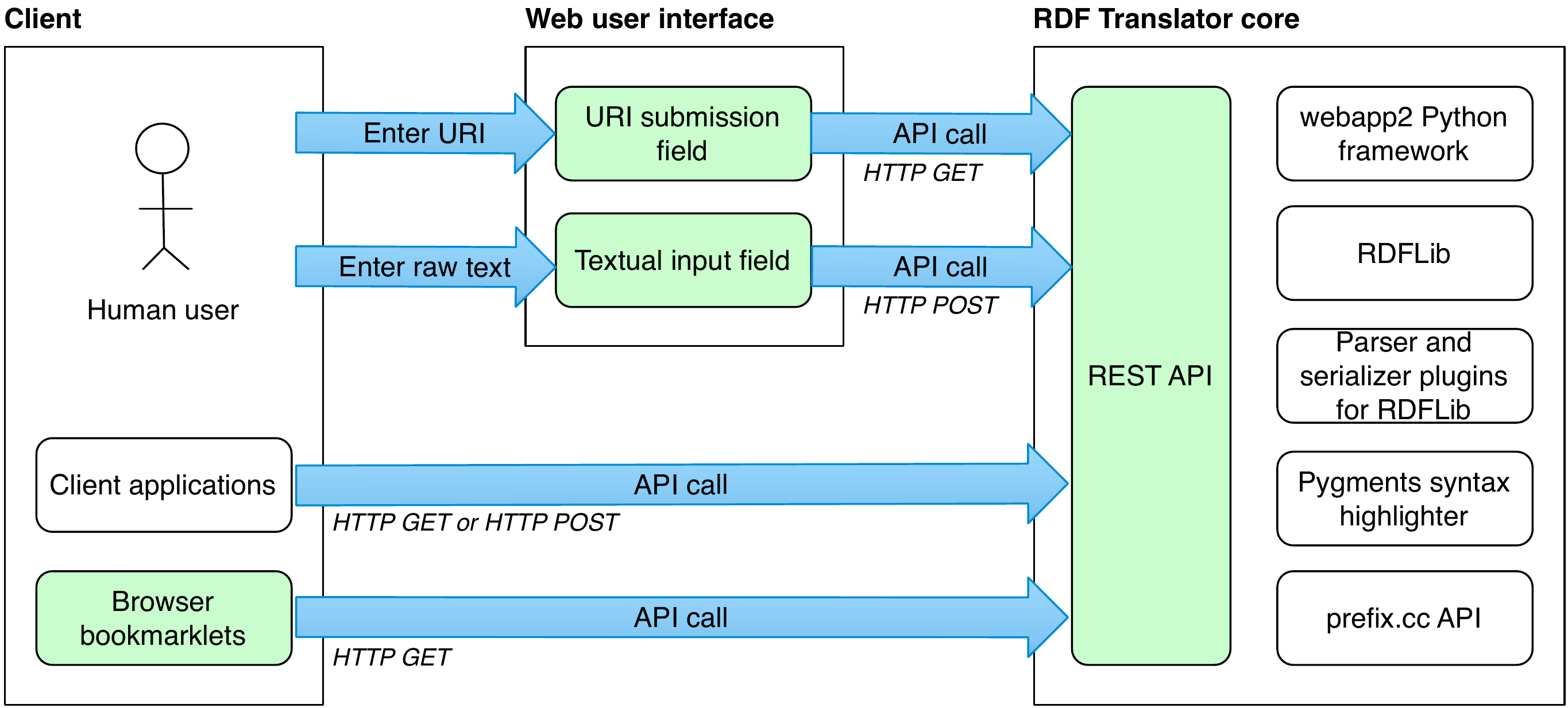}
\caption{Service architecture of \textit{RDF Translator}}
\label{fig:rdftranslator-architecture}
\end{figure}

Fig. \ref{fig:rdftranslator-architecture} depicts the architecture of RDF Translator. At the core of the Web service is a powerful REST API that handles all incoming requests, be it direct requests from HTTP-capable client applications, browser bookmarklets, or indirect requests via the Web user interface. The REST API supports both HTTP GET and HTTP POST request methods, whereby the latter is more suitable than the former for submitting raw data to be converted by the service. For human users the Web service offers an intuitive Web user interface, which provides two options, namely the provision of a Web resource URI which contents are to be converted, or raw textual input to be translated into a target format. More details about the Web user interface follow in Section \ref{subsec:webuserinterface}.

The converter is implemented as a Google AppEngine Web service (based on the \textit{webapp2} Python framework) and can thus scale up to many parallel requests given that Google allows for paid plans with quite flexible resource allocation. The service takes advantage of the RDFLib library\footnote{\url{http://www.rdflib.net/}} for parsing, manipulating and serializing RDF graphs. With respect to covering the full conversion cycle, RDFLib in its current version is missing serializers for RDFa and Microdata. Initially we tried to fill this gap by using RDF2RDFa and RDF2Microdata (see Section \ref{sec:relatedwork}), however, their outputs are unsatisfying because maintenance of these services has stopped long time ago, so we decided to provide two custom plugins compatible with RDFLib that we made publicly available online\footnote{\url{https://github.com/alexstolz/rdflib-rdfa-serializer} and \url{https://github.com/alexstolz/rdflib-microdata-serializer}}.
The full source code of the Web service has been published as a Bitbucket repository\footnote{\url{https://bitbucket.org/alexstolz/rdf-translator}}. It is licensed under a LGPL license and people are invited to run their own service instances on Google AppEngine.

\subsection{Web User Interface}
\label{subsec:webuserinterface}

\begin{figure}[t]
\center
\includegraphics[width=\textwidth]{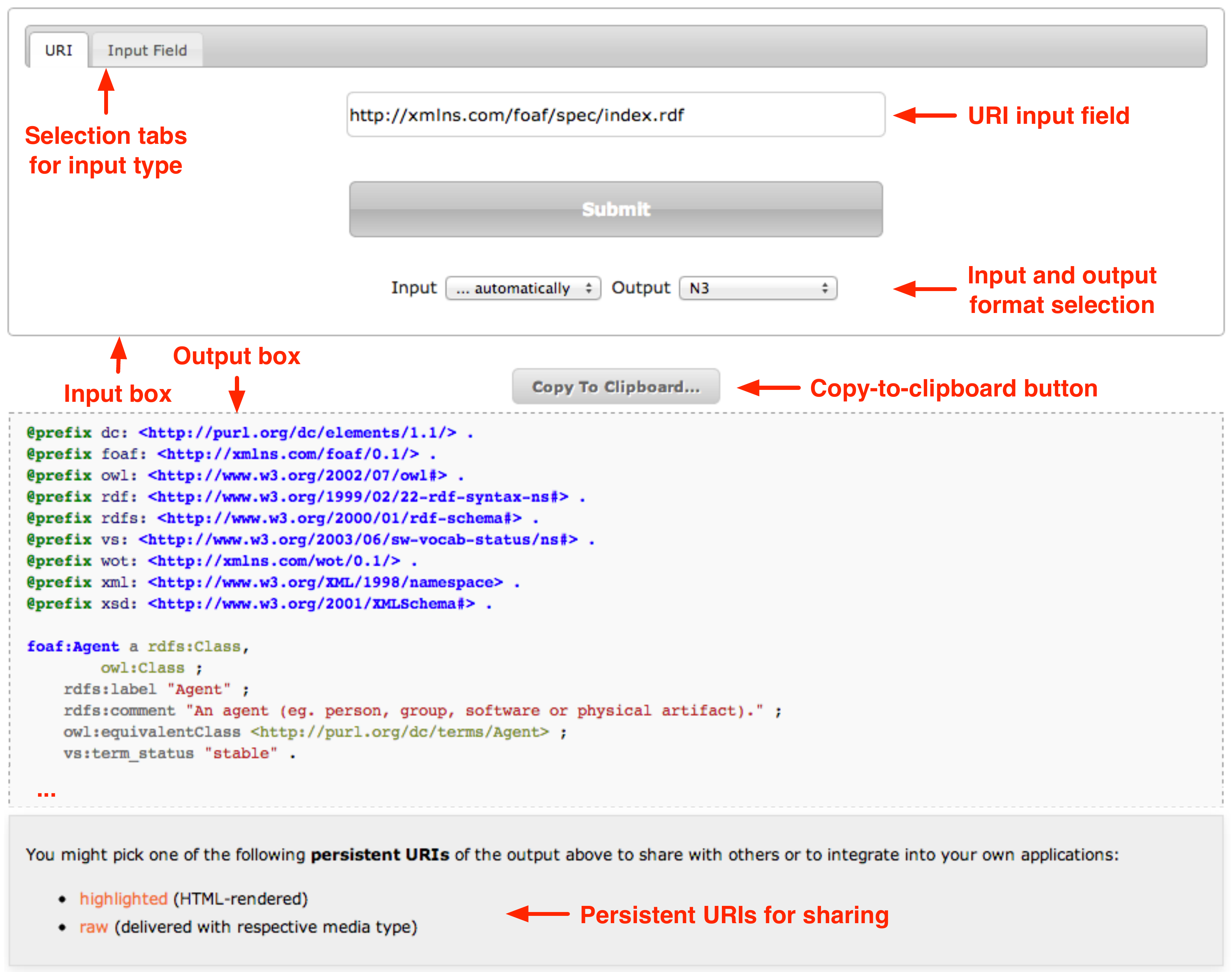}
\caption{Screenshot of the \textit{RDF Translator} service}
\label{fig:rdftranslator-screenshot}
\end{figure}

The user interface was developed with user-friendliness in mind, thus kept clean and straightforward. 
Fig. \ref{fig:rdftranslator-screenshot} illustrates a screenshot of the Web front end of the Web service, sketching the input box, the syntax-highlighted output and the copy-to-clipboard button to ease copy-and-paste.
It further can be seen that the input box consists of tabs, one for the URI submission (default option) and the other one for the textual input. In the given screenshot the URI tab is activated, thus showing a URI input field along with the submit button and selectors for the input and output formats.
The data formats currently supported by the Web service are:
\begin{itemize}
\item RDFa,
\item Microdata,
\item RDF/XML (standard syntax, plus a more concise XML syntax, i.e. replacing \textit{rdf:Description} nodes by typed nodes),
\item N3,
\item N-Triples,
\item RDF-JSON (standard syntax, plus a more concise JSON syntax, i.e. using namespace prefixes instead of full URIs), and
\item JSON-LD.
\end{itemize}

Furthermore, the input format can be determined automatically by means of content negotiation, under the premise that the correct media type was supplied. Otherwise, the automatic detection of the document format fails with an error message.

The Web front end consists of a HTML5 document with JavaScript (JQuery) for user interaction, asynchronous communication, and manipulation of the document tree.

\subsubsection{URI Submission.}

The main element for the URI submission is the input text field for entering the URI.
As long as there is input missing or the supplied URI is invalid, the URI input field remains red-shaded. This is a feature supported by HTML5 which attracts attention of the user to prevent him from submitting incorrect data. However, for simple HTTP addresses the \textit{http://}-prefix need not be explicitly supplied with the URI. According to this, it is possible to shorten \url{http://www.example.com} with \url{www.example.com}, but not \url{https://www.example.com}. For helping the user to get started with the Web service, the URI input field is pre-filled with a Web page URI that contains semantic descriptions.

The Web service provides key shortcuts for faster form submissions.
If the page focus is in the URI input field or in the input or output selectors, a simple keystroke (pressing the \textit{return}-key) suffices to trigger the conversion process. So there is actually no need to press the \textit{Submit}-button, which may help speed up repetitive tasks like the tedious task of translating a moderate list of URIs into other formats.

For sharing results with other people the Web service returns persistent URIs together with the conversion output. Unlike as for the textual input mode, for URI submissions there appears an additional grey box below the converter output. It contains two persistent links for easy sharing with other developers, namely one link that refers to the syntax-highlighted version of the output for sharing between humans, and a second link that points to the raw output format with the proper media type supplied, hence being suitable for integration with other Web applications.

\subsubsection{Textual Input.}

The textual input allows developers to enter raw input data into a text area. Since users submit plain data, the automatic format detection does not work for this type of transformation unless the format is guessed by chance. In other words, content negotiation fails because there is no metadata available as when calling a URI that returns HTTP response headers with content type declarations supplied.

Developers are again actively supported by being offered working examples for every type of serialization format to become acquainted more easily with the Web service.

\subsubsection{Syntax Highlighting.}

Syntax highlighting is a visual means that most developers using advanced text editors to write and read code are familiar with. Similar to programming languages, also the different syntax variants for RDF can be made better readable for humans by highlighting terms according to grammar and keywords. Our Web service applies syntax highlighting to the output of the translated contents in the Web interface or, if requested explicitly, using the REST API.

The Web service accomplishes syntax highlighting by virtue of the sophisticated \textit{Pygments}\footnote{\url{http://pygments.org/}} library for Python. \textit{Pygments} supports all required syntaxes, i.e.
RDFa and Microdata using the HTML formatter;
RDF/XML taking advantage of the XML formatter;
RDF/JSON and JSON-LD by use of the JSON lexer; and
N3 and N-Triples by means of a custom-built Notation 3 lexer.

\subsubsection{Bookmarklets.}

For developers that do not prefer to visit the Web service every time they find an interesting Web resource with embedded RDFa or Microdata, bookmarklets are a convenient means for quickly extracting the contained structured markup of any Web page and presenting it in the developer's preferred syntax. Bookmarklets are JavaScript snippets suitable for being added to bookmark folders and toolbars of modern browsers.
This way a simple click on the created bookmark is sufficient to translate the contents of the current Web resource into an arbitrary target RDF syntax. The bookmarklet code used to convert structured data in the form of RDFa to syntax-highlighted Notation 3 code looks as follows:
\begin{lstlisting}[language=javascript]
javascript:location.href='http://rdf-translator.appspot.com/convert/rdfa/n3/html/'+encodeURIComponent(location.href);
\end{lstlisting}
On the landing page of our Web service we are providing a comprehensive matrix table with all possible combinations of syntax conversions.

\subsubsection{Other Features.}

In addition to the features presented so far, the Web user interface can further assist human users in various aspects. This includes:
\begin{itemize}
\item copy-to-clipboard functionality that saves the effort for selecting and copying the output text to the system clipboard (see Fig. \ref{fig:rdftranslator-screenshot}),
\item feedback link (not visible in Fig. \ref{fig:rdftranslator-screenshot}) for contacting the developers, and
\item verbose error messages that can provide helpful explanations about failed translations.
\end{itemize}

\subsection{REST API}

As mentioned previously, our Web service is based on the REST architecture that was first introduced by R. T. Fielding in his PhD thesis in 2000 \cite{Fielding2000}.
The core aspects of REST are the client-server architecture, statelessness of client and server across multiple requests, uniform identifiers for addressing resources with specific representations, and self-descriptive messages. This way REST scales very well, making it a best practice for Web services design and at the same moment a recommended architectural style for the World Wide Web.
As such a RESTful Web service not only permits to be used through a Web front end, but also by other client applications that make use of the Hypertext Transfer Protocol (HTTP) \cite{Fielding1999} and related features like content negotiation, e.g. by command-line tools like \textit{curl}\footnote{\url{http://curl.haxx.se/}} or even browser plugins (and bookmarklets).

We address the requirements of the REST architecture by providing cool URIs, HTTP GET and HTTP POST methods, and HTTP content negotiation where requested media types define how to represent a resource.

\subsubsection{Cool URIs.}

In 1998, Tim Berners-Lee published an article \cite{Berners-Lee1998} where he postulated the use of cool URIs on the Web. The idea is to leave out details in the URI string that are subject to likely change in the future, e.g. status information about the document (e.g. \textit{draft}, \textit{final}), underlying software mechanisms (e.g. \textit{.php}, \textit{cgi-bin}), or even metadata (e.g. author information, or storage details like disk names). Otherwise it involves additional maintenance effort (e.g. setting up appropriate redirects), always at the risk of loosing users or breaking up with applications that are taking advantage of the resource. A well-planned and accurately organized URI design could help mitigate these problems. URIs qualify as cool URIs as soon as they are simple, stable and manageable \cite{Sauermann2008}.

For our REST API we decided to encode variables in the hierarchical part of the URI scheme, and thus not to attach query string parameters to the converter script.
The respective URI for the conversion to a raw target format then looks as given by the following URI pattern:
\begin{lstlisting}
http://rdf-translator.appspot.com/convert/<source>/<target>/<uri>
\end{lstlisting}
The placeholder \textit{source} within angle brackets denotes the input format, \textit{target} the output format, and \textit{uri} the input URI, respectively.
The values eligible for the source format are \textit{rdfa}, \textit{microdata}, \textit{xml}, \textit{n3}, \textit{nt}, \textit{rdf-json}, and \textit{json-ld}. For the specification of the target format the service understands two additional values, namely \textit{pretty-xml} for concise RDF/XML, and \textit{rdf-json-pretty} for concise RDF/JSON, respectively. 
For the input option, there is the possibility to let the service automatically detect the media type of the input resource. This is achieved by supplying the value \textit{detect}.

For the human-readable version, i.e. the syntax-highlighted output obtained using proper HTML and CSS formatting, the pattern is slightly different (note the additional substring \textit{/html} in the URI pattern):
\begin{lstlisting}
http://rdf-translator.appspot.com/convert/<source>/<target>/html/<uri>
\end{lstlisting}


The same URI pattern applies to conversions that send the data in the message body of the request, i.e. HTTP POST methods. Instead of providing an input URI, the string \textit{content} is to be supplied along with a key-value pair \verb|content=<data>| in the message body, where \textit{data} is the URL-encoded input data to be translated.

\subsubsection{Content Negotiation.}

The REST API internally implements HTTP content negotiation (\cite{Fielding1999}, Section 12) and thereby is able to request document representations according to their media types and to return syntax conversions with proper content types.

Given that Web servers hosting the input URI support content negotiation, the API can take advantage of HTTP \textit{Accept} headers (\cite{Fielding1999}, Section 14.1), a method to express the preferred content type in order to request a specific document representation.
Similarly, our API returns the corresponding media types together with the RDF serializations that were specified by the URI pattern when invoking the service API.
Table \ref{tab:content-type-document-format-mapping} gives a mapping of the media types that apply to the respective serialization formats requested by using the output field supplied with the URI.
For the syntax-highlighted (``pygmentized'') output aimed at being viewed in the Web browser, the returned media type gives constantly \textit{text/html}, no matter what the target format in the URI was set to.

\begin{table}
\caption{Content types returned for serialization formats}
\label{tab:content-type-document-format-mapping}
\center
\begin{tabular}{|l|l|l|}
\hline
\textbf{Document format} & \textbf{Output value in URI} & \textbf{Content type}\\\hline
\textit{HTML (pygmentized)} & exception: \textit{/html} & text/html\\
RDFa & rdfa & text/html\\
Microdata & microdata & text/html\\
RDF/XML and concise RDF/XML & xml, pretty-xml & application/rdf+xml\\
Notation 3 & n3 & text/n3\\
N-Triples & nt & text/plain\\
RDF/JSON and concise RDF/JSON & rdf-json, rdf-json-pretty & application/json\\
JSON-LD & json-ld & application/json\\\hline
\end{tabular}
\end{table}

\subsubsection{GET Request.}

We demonstrate the use of HTTP GET requests by giving a concrete example using the command-line tool \textit{curl} and translating the RDF/XML representation of the FOAF vocabulary specification to Notation 3:
\begin{lstlisting}
curl "http://rdf-translator.appspot.com/convert/xml/n3/http://xmlns.com/foaf/spec/index.rdf"
\end{lstlisting}

Internally, the Web service invokes the input URI requesting RDF/XML and converts the retrieved contents into the target format, namely Notation 3.
By default, \textit{curl} uses the HTTP GET method to retrieve content from URIs.

\subsubsection{POST Request.}

In addition to HTTP GET requests, the converter permits to perform HTTP POST requests with the data attached in the message body of the request.

In order to translate raw data using the command line, it is just sufficient to invoke the following \textit{curl} command, supplying the option \textit{--data-urlencode} that denotes a HTTP POST request:
\begin{lstlisting}
curl --data-urlencode content="@prefix : <http://example.org/#> . :a :b :c ." http://rdf-translator.appspot.com/convert/n3/nt/content
\end{lstlisting}
The response body of the execution of the last command results in the following output in N-Triples format:
\begin{lstlisting}
<http://example.org/#a> <http://example.org/#b> <http://example.org/#c> .
\end{lstlisting}

Alternatively, we could have achieved the same by using a local file containing the equivalent Notation 3 snippet as in the \textit{curl} command before. The following command translates the contents of a local file to N-Triples and saves the results as a new file:
\begin{lstlisting}
curl --data-urlencode content@example.n3 http://rdf-translator.appspot.com/convert/n3/nt/content > example.nt
\end{lstlisting}

\subsection{Cross-Origin Resource Sharing Support}

Nowadays, many applications built for the Web provide user interaction without having to leave the page for every single action the user performs. This Web page dynamics is enabled by JavaScript and its \textit{XMLHTTPRequest} object to retrieve data at a given URL, and supported by almost every modern Web browser. However, in Web browsers there are security constraints in place (the ``same origin policy''\footnote{\url{http://www.w3.org/Security/wiki/Same_Origin_Policy}}) that prevent JavaScript applications from loading content from external resources that are not under the control of the Web site.

Cross-origin resource sharing \cite{vankesteren2013} tackles the problem of the same origin policy by adding a HTTP response header to allow open access across domain boundaries. The header field looks as follows and is included in the response header of every response message returned by our REST API:
\begin{lstlisting}
Access-Control-Allow-Origin: *
\end{lstlisting}

\subsection{Vocabulary Detection}

\textit{RDFLib}, the RDF library that our service relies on, already defines a considerable number of namespace prefixes of popular vocabularies. They are used to add generally accepted namespace prefix declarations to syntaxes like Notation~3.
Unfortunately, prefix declarations of specific or newly created vocabularies are not found in this seed list.
To provide meaningful prefixes for unknown vocabularies, though, our service integrates the prefix.cc\footnote{\url{http://www.prefix.cc/}} lookup service as a fallback solution, whereby it applies the reverse lookup feature of the service in order to obtain the appropriate namespace prefix. In particular, the service first checks if a vocabulary is already in the list of vocabularies known to RDFLib. If this is not the case, it claims the prefix of the vocabulary URI from prefix.cc.
The nice side effect of this solution is that someone creating a new ontology can register the vocabulary prefix at prefix.cc and afterwards will immediately be returned the correct prefix name in the RDF Translator.


\section{Use Cases} 
\label{sec:usecases}

The typical Semantic Web development process involves many requirements that are insufficiently addressed by many of the tools presented in Section \ref{sec:relatedwork}.
Such requirements range from technical solutions to active user support during the development and testing of Semantic Web artifacts.
The technical aspect involves providing an otherwise unsupported document format in the syntax preferred by a consuming application.
Similarly, in collaborative tasks it often proves useful to be able to share human-readable representations of data with others, such as exchanging Notation 3 code instead of a Web document with RDFa in HTML.

Based on this intuition, we defined two requirements that we successfully applied in real use cases and that we think many of the other existing Web services for RDF syntax conversions are not able to address well.

\begin{requirement}
[REST API]
Add the ability to serve arbitrary RDF output formats (N3, RDF/XML, N-Triples, etc.) for an ontology document based on a specific RDF input format (e.g. N3).
\end{requirement}

Ontologists create their Web ontologies either by using one of the existing ontology editors or, if the ontologies are rather small, by hand-crafting them using Notation 3. Afterwards they locally convert the documents to other popular document formats and publish them altogether on the Web. This is at the cost of having to repeat this task on every update of the ontology. It would be more convenient to be able to publish the document in the preferred syntax on the Web and then to configure the Web server (e.g. Apache) to serve the ontology in different formats, taking advantage of RDF Translator.

The Exchange Rate Ontology\footnote{\url{http://purl.org/xro/}} is configured exactly this way. An example to get the ontology description in N-Triples syntax taken from the Apache \textit{.htaccess} configuration file looks as follows:
\begin{lstlisting}
RewriteCond %{HTTP_ACCEPT} text/plain
RewriteRule ^ns$ http://rdf-translator.appspot.com/convert/n3/nt/http://www.stalsoft.com/ontologies/xro/ns.n3 [P,L]

RewriteRule \.nt$ http://rdf-translator.appspot.com/convert/n3/nt/http://www.stalsoft.com/ontologies/xro/ns.n3 [P,L]
\end{lstlisting}

This way the ontology would return N-Triples either if the client requests \textit{text/plain} (first rewrite rule in the previous code snippet) or by using the file extension \textit{.nt} (second rewrite rule).

\begin{requirement}
[Syntax highlighting]
Translate a RDF/XML ontology document to syntax-highlighted Notation 3 for viewing in the Web browser.
\end{requirement}

RDF Translator supports raw output delivered with the respective media types and HTML-rendered output with syntax highlighting. The latter is achieved by including \textit{/html} in the hierarchical part of the URI in front of the specification of the input URI or, for POST requests, right before the \textit{content} value in the URI string. Let us assume that we would like to show the contents of the Exchange Rate Ontology in human-readable Notation 3 with syntax highlighting enabled (Fig. \ref{fig:usecase2}).
The corresponding URI is:
\begin{lstlisting}
http://rdf-translator.appspot.com/convert/xml/n3/html/http://purl.org/xro/ns
\end{lstlisting}

\begin{figure}[t]
\center
\includegraphics[width=\textwidth]{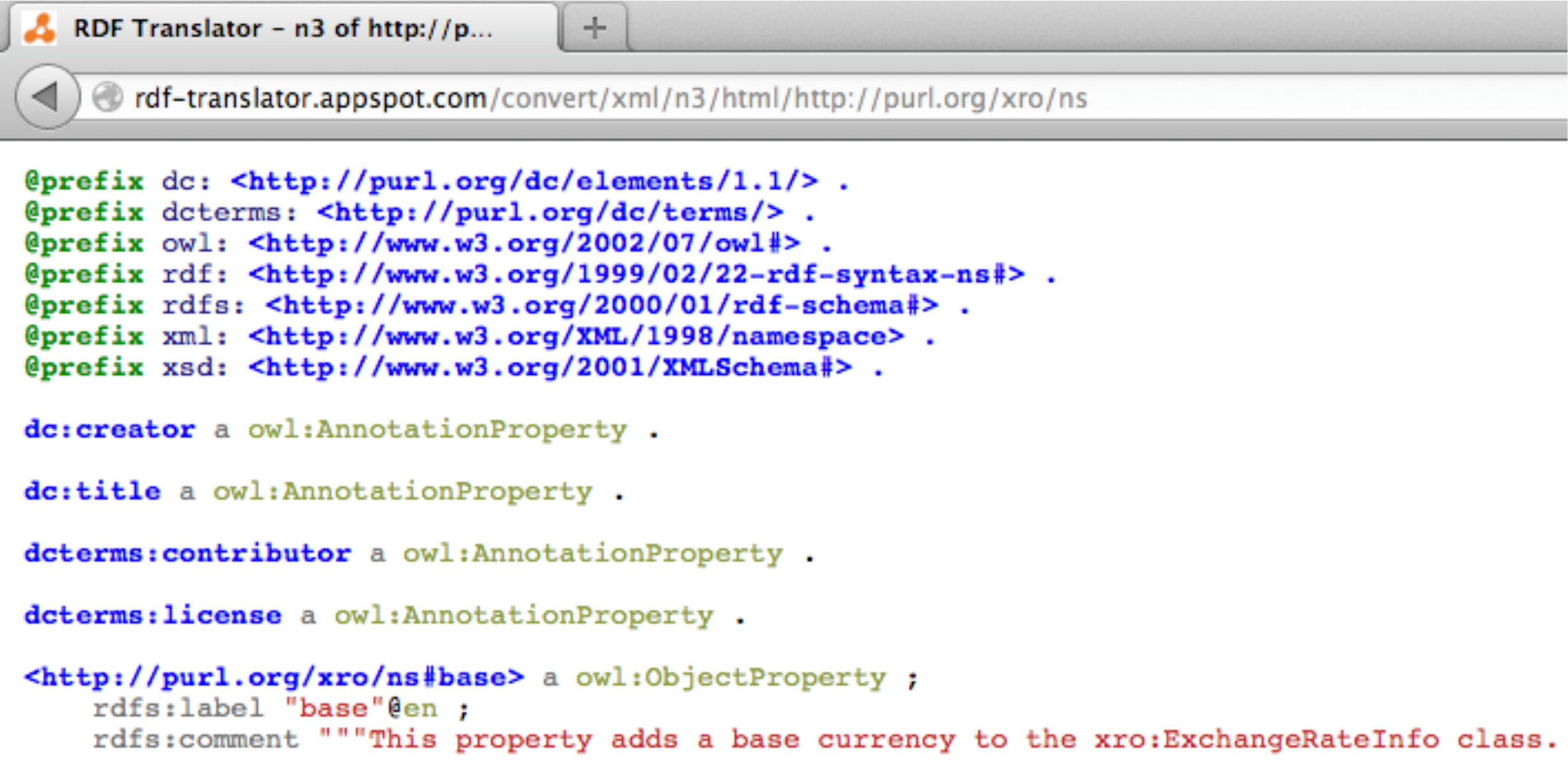}
\caption{Syntax-highlighted N3 snippet in Web browser}
\label{fig:usecase2}
\end{figure}

There are three important technical details in this example that need further explanation:
First, RDF Translator follows HTTP redirect links, because the short URI \verb|http://purl.org/xro/ns| links to the actual location of the deployment at \verb|http://www.stalsoft.com/ontologies/xro/ns|.
Second, it requests the RDF/XML representation of the document using HTTP content negotiation.
And finally, it translates the contents to Notation 3 and, using Pygments, wraps them into syntax-highlighted HTML code for viewing in the Web browser.

The bookmarklet feature presented in Section \ref{sec:approach} is taking advantage of this functionality, i.e. there is one among the 72 bookmarklets that is able to show markup encoded as RDF/XML in syntax-highlighted Notation 3.
Furthermore, this functionality has also been integrated into a browser extension for Google Chrome, the Grome extension\footnote{\url{http://www.stalsoft.com/grome}} that detects GoodRelations markup, either as RDFa or Microdata, in Web pages and allows to quickly show the embedded metadata in the preferred syntax.


\section{Conclusions}
\label{sec:conclusions}

In this paper we presented RDF Translator, a RESTful Web service that comes with a developer-friendly user interface and a REST API. The Web service is intended as a comprehensive solution for translating between the most popular serialization formats currently available on the Semantic Web. Our proposal does not only focus on the technical aspects of supporting the development of Semantic Web applications with syntax transformation capability, but also on collaborative aspects of the development process.
What makes our solution unique among the Web services presented in Section \ref{sec:relatedwork} is the combination of features like the bidirectional conversion capability of a number of popular RDF data formats, the possibility of syntax highlighting for all of the supported serialization formats, link sharing functionality for better collaboration, a clean and straightforward Web user interface with active user support, and compliance with cutting-edge Web technologies.

We frequently update the libraries that underlie our Web service to the most recent versions, thus we think we can reflect well the latest changes to the different formats. As such we think our service represents a well-maintained, authoritative conversion service on the Web that can be a useful tool for a large number of Semantic Web developers.

As future work we consider to improve error detection and verbosity of our Web service in terms of providing status codes and corresponding error messages like the \textit{Any23} Web service is already doing. This would improve interoperability with Web applications that take advantage of RDF Translator and at the same time it would allow developers to provide us with more accurate feedback.


\subsubsection*{Acknowledgments.} 
The work described in this paper has been partly supported
by the German Federal Ministry of Education and Research (BMBF) by a grant under the KMU Innovativ program as part of the Intelligent Match project (FKZ 01IS10022B); and
by the EUREKA and European Commission under the Eurostars program as part of the Ontology-based Product Data Management (OPDM) project (FKZ 01QE1113D).

\bibliographystyle{splncs03}
\bibliography{dissertation-publications-rdftranslator}

\end{document}